\documentclass[twocolumn,superscriptaddress,amsmath,amssymb]{revtex4}
\usepackage{graphicx}
\usepackage{dcolumn,xcolor}
\usepackage{amsmath} 
\usepackage{amssymb}
\usepackage{amsfonts}
\usepackage{bm}

\usepackage[latin1]{inputenc}

\makeatletter
\newcommand\xleftrightarrow[2][]{%
  \ext@arrow 9999{\longleftrightarrowfill@}{#1}{#2}}
\newcommand\longleftrightarrowfill@{%
  \arrowfill@\leftarrow\relbar\rightarrow}
\makeatother

\begin{document}



\title{The size of the Sun}

\author{M.A. Fardin}
\altaffiliation[Corresponding author ]{}
\email{marc-antoine.fardin@ijm.fr}
\affiliation{Universit\'{e} de Paris, CNRS, Institut Jacques Monod, F-75013 Paris, France}
\affiliation{The Academy of Bradylogists, 75013 Paris, France}
\author{M.~Hautefeuille}
\affiliation{Institut de Biologie Paris Seine, Sorbonne Universit\'{e}, 7 quai Saint Bernard, 75005 Paris, France}

\date{\today}

\begin{abstract}
Why does the Sun have a radius around 696000~km? We will see in this article that dimensional arguments can be used to understand the size of the Sun and of a few other things along the way. These arguments are not new and can be found scattered in textbooks. They are presented here in a succinct way in order to better confront the kinematic and mechanical viewpoints on size. We derive and compare a number of expressions for the size of the Sun and relate large and small scales. We hope that such presentation will be useful to students, instructors and researchers alike.  
\end{abstract}

\maketitle

In \textit{The Character of Physical Law} Richard Feynman stated that ``every theoretical physicist who is any good knows six or seven different theoretical representations for exactly the same physics. He knows that they are all equivalent, and that nobody is ever going to be able to decide which one is right at that level, but he keeps them in his head, hoping that they will give him different ideas for guessing.''~\cite{Feynman2017} Following Feynman's advice, we provide several different expressions for the radius of the Sun. More generally, this will lead us to expressions for the size of other stars and astronomical bodies like planets and satellites, and eventually to a discussion of size in a broader sense. 

Many excellent textbooks review the multiple sciences involved in the understanding of stars, their mechanics, thermodynamics, acoustics, the intricacies of radiative and convective processes, the interplay of atomic and nuclear physics, or the crossover between quantum and relativistic phenomena. To assist writing this article, we principally used the book \textit{Stellar Structure and Evolution}~\cite{Kippenhahn1990}. These many approaches cover widely varying scales and viewpoints, and they can be followed quite rigorously in many cases, yielding very good agreement with {observations}. The present article proposes a walk through some of the most thought-provoking formulas offered by this field in order to show how the different viewpoints are related, painting a kaleidoscopic answer to the question: `what is the size of the Sun?'  

Volumes have been written on the physics of stars because each argument that we will highlight can actually be pushed to a high degree of precision through adequate consideration of the geometric and dynamical subtleties underlying them. To enable a wide-ranging exploration in the short span of this article, we will have to neglect these subtleties; we will rely mostly on dimensional arguments. Formulas that will be presented have to be understood as approximate, neglecting small numerical factors of order $1$. For instance, we will say that the volume of a sphere of radius $R$ is $R^3$, omitting the factor $4\pi/3$. Similarly, we will neglect careful integration of spatially varying fields like pressure and density and only rely on approximate bulk averages. At no point shall we venture to state exact results, so we have used the sign `=' to stand for approximate equality, where others usually use `$\propto$' or `$\simeq$'. We have reserved these last two symbols to respectively stand for scaling relations and approximate numerical results. Thus, when we will say that $A\propto B$, we mean that $A$ is proportional to $B$, and when we say that $A\simeq a u_a$, we mean that the value of $A$ in units $u_a$ is approximately the number $a$.   

\subsection*{Hydrostatic equilibrium}
The most common way to approach the size of the Sun and similar stars is to consider that it derives from an equilibrium between two {competing factors}. The first {factor} is gravity, which tends to compress the Sun. It can be expressed as a force per unit volume $\Psi$ (dimensions $\mathcal{M} \mathcal{L}^{-2} \mathcal{T}^{-2}$). The second {factor} is the pressure inside the sun, resisting further compression. It can be expressed as a force per unit area $\Sigma$ ($\mathcal{M}\mathcal{L}^{-1}\mathcal{T}^{-2}$). From these two quantities, dimensions can combine to produce a length: 
\begin{equation}
R=\frac{\Sigma}{\Psi}
\label{hydro}
\end{equation}
This length $R$ is the radius of the Sun, understood as a ratio of pressure and force density. This approach is generally called that of `hydrostatic equilibrium', because it basically has the same form as the relation between the pressure inside a fluid and its own weight~\cite{Landau2013}. This argument of hydrostatic equilibrium first emerged in the late 19th century in studies by Lane, Kelvin and Helmholtz~\cite{Powell1988,Arny1990}. The equilibrium can also be written as $\Psi=\Sigma/R$, which relates the force density to a pressure gradient, or as $\Psi R=\Sigma$, which relates the gravitational inward pressure on the left to the outward pressure on the right. The radius can also be seen as the result of the balance of forces $\Psi R^3 = \Sigma R^2$. Of course, neither pressure nor force density are homogeneous inside the Sun, so the quantities $\Psi$ and $\Sigma$ should be understood as giving average orders of magnitude.  

To evaluate the relevance of Eq.~\ref{hydro}, we can express the force density as the product between the average density of the Sun and the acceleration of gravity on its surface: $\Psi=\rho g$. Using the experimental values $\rho \simeq 10^3$~kg/m$^3$ and  $g\simeq 3~10^2$~m/s$^2$, this gives $\Psi \simeq 3~10^5$~N/m$^3$~\cite{Stix2012}. Since the observed radius of the Sun is $R\simeq 7~10^8$~m, for Eq.~\ref{hydro} to be valid would imply that the average pressure of the Sun is  $\Sigma\simeq 2~10^{14}$~Pa. In practice, the pressure of the Sun varies from about $10^{-2}$~Pa in its corona~\cite{Gomez2019}, to $3~10^{16}$~Pa in the core, so $\Sigma\simeq 2~10^{14}$~Pa must be understood as a bulk average~\cite{Stix2012}. 

\subsection*{Sound speed and gravitational acceleration}
The hydrostatic equilibrium as expressed in Eq.~\ref{hydro} is the archetype of a mechanical expression for a length scale. By this, we mean that the radius $R$, with dimension $\mathcal{L}$, is expressed as a ratio between two `mass-carrying quantities', i.e. quantities with mass in their units. In the case where the force density can be expressed as a weight density $\Psi=\rho g$, the hydrostatic equilibrium can also be turned into a purely kinematic expression: 
\begin{equation}
R=\frac{v_s^2}{g}
\label{gvs}
\end{equation}
 \noindent where $v_s=(\Sigma/\rho)^\frac{1}{2}$ is the `speed of sound' inside the Sun, i.e. the speed of mechanical waves~\cite{Landau2013}. In this expression of the radius, the terms of the ratio on the right-hand side do not have any mass in their units. The dimensions of the speed $v_s$ are  $\mathcal{L}\mathcal{T}^{-1}$, and the dimensions of the acceleration $g$ are $\mathcal{L}\mathcal{T}^{-2}$. Overall, the ratio $v_s^2/g$ has the dimensions of a distance. Here again, the terms of the equation can be moved around so as to gain new insight. For instance, one can notice that $R/v_s$ is the time required for sound waves to travel across the Sun. According to Eq.~\ref{gvs}, this time scale should be equal to $v_s/g$. We will shortly see how to interpret this time scale. 
 
If the radius and accelerations are known, Eq.~\ref{gvs} can be used to estimate the average sound speed inside the Sun, as $v_s = (Rg)^\frac{1}{2}\simeq 4~10^5$~m/s. This value is indeed the right order of magnitude of the bulk average~\cite{Stix2012,Christensen2002}.    
 
 \subsection*{The gravitational constants}
We mentioned that at the surface of the Sun the gravitational acceleration is $g\simeq 275$~m/s$^2$, yet on Earth it is $g\simeq 9.8$~m/s$^2$. This difference is due to the fact that behind the acceleration $g$ hides the mass and the radius $R$ itself. Indeed, the weight $F_i=m_i g$ of an object of mass $m_i$ at the surface of the Sun can be expressed more generally from Newton's formula as $F_i=GM m_i/R^2$, where $G\simeq 6~10^{-11}$~kg$^{-1}$ m$^3$ s$^{-2}$ is the universal gravitational constant, and where $M\simeq 2~10^{30}$~kg is the mass of the Sun. Then, we can replace $g$ by $G M/R^2$ and express the density of the Sun as $\rho=M/R^3$. By inserting these expressions for $g$ and $\rho$ into Eq.~\ref{gvs} and solving for $R$ we get: 
\begin{equation}
R= \Big(\frac{GM^2}{\Sigma}\Big)^\frac{1}{4}
\label{Geq1}
\end{equation}
 If instead we express the mass {in terms of} the density, we get: 
\begin{equation}
R= \Big(\frac{\Sigma}{G \rho^2} \Big)^\frac{1}{2}
\label{Geq2}
\end{equation}
These two equations are mechanical in the sense defined above: they involve quantities with mass in their dimensions. However, in contrast to Eq.~\ref{hydro} these equations involve three rather than two quantities. 

\subsection*{Sound/escape speed and free-fall time}
To a gravitational field with acceleration $g=G M/R^2$ one can also associate an `escape speed' $v_e=(GM/R)^\frac{1}{2}$, which gives the typical speed necessary to escape the attraction of a body of mass $M$ and radius $R$. Coincidentally, this speed is also the orbital speed at a distance $R$. Either way, $v_e$ is the scale of speed set by gravity. One can see that the escape speed can be expressed as $v_e=(gR)^\frac{1}{2}$, which is also the expression of the sound speed deduced from Eq.~\ref{gvs}. Hence, the radius of the sun can be understood as corresponding to the {identity} between the sound speed and the escape speed, $v_e=v_s$:
\begin{equation}
\Big(\frac{G M}{R}\Big)^\frac{1}{2} = \Big(\frac{\Sigma}{\rho}\Big)^\frac{1}{2}
\end{equation}
This approach is usually associated with Jeans, {who} derived the condition for a gas cloud to collapse into a star to be $v_e\gtrsim v_s$~\cite{Jeans1902}. 

Now that we know that for a gravity-bound body at hydrostatic equilibrium the sound speed is equal to the escape speed, we can give an interpretation to the time scale $v_s/g$. Indeed, since $v_s=v_e=(gR)^\frac{1}{2}$, then we can define a time scale $\tau=v_s/g=(R/g)^\frac{1}{2}=(G\rho)^{-\frac{1}{2}}$. This time scale is often called the `free-fall time' and corresponds to the time that a body would take to collapse under its own gravitational attraction~\cite{Kippenhahn1990}. For the Sun, $\tau\simeq 1$~h. This time scale does not depend on the absolute size of the object but on its density. Since the Sun and a human being have similar densities, they would collapse after a similar time of one hour. 

The radius $R$ can be understood as the distance traveled by sound over a time equal to $\tau$. Generally, we have
\begin{equation}
{R= v_s \tau = v_e \tau}
\end{equation}

\subsection*{Schwarzschild radius}
Gravitation associates a {special} radius to any object with a given mass. This is the Schwarzschild radius~\cite{Schwarzschild1916}, which corresponds to the size of a black hole with that mass. For the Sun, this radius is $R_s=GM/c^2 \simeq 1$~km, where $c\simeq 3~10^8$~m/s is the speed of light. We can use Eq.~\ref{Geq1} to express the size of the Sun {in terms of} its Schwarzschild radius as:
\begin{equation}
R=R_s \frac{c^2}{(v_e^3 v_s)^\frac{1}{2}} =R_s \Big(\frac{c}{v_e}\Big)^2 = R_s \Big(\frac{c}{v_s}\Big)^2
\label{Blackhole}
\end{equation}
In this {identity}, the dimension of length is provided by $R_s$, and the rest of the equation is a dimensionless ratio built out of the speeds $v_s$, $v_e$ and $c$. In the limit where $v_e=v_s=c$, we have $R=R_s$, i.e. if the escape/sound speed is equal to the speed of light, then the {object} is a black hole. 

\subsection*{Solar internal energy}
So far, we expanded on Eq.~\ref{hydro} by specifying the content of the force density $\Psi=\rho g=\rho^2 GR=GM^2/R^5$ when the inward force comes from gravity. In contrast, we barely investigated the pressure $\Sigma$; we just related it to the sound speed. Density is quite naturally connected to a ratio between mass and volume, as in $\rho=M/R^3$. In a similar way, the dimensions of pressure suggest that it can be expressed as a density of energy since $\mathcal{M}\mathcal{L}^{-1}\mathcal{T}^{-2}$=$\mathcal{M}\mathcal{L}^{2}\mathcal{T}^{-2} / \mathcal{L}^{3} $. Thus, one can define a solar energy as $E=\Sigma R^3$ and express the radius of the Sun as: 
\begin{equation}
R=\frac{E}{F}=\frac{G M^2}{E}\label{solarE}
\end{equation}
\noindent where $F=\Psi R^3=GM^2/R^2$ is the scale of the self gravitational force. The second expression will later be useful. The first equation can also be rearranged to express the solar energy as $E=M v_e^2 = M g R\simeq 10^{42}$~J. As a comparison, Type Ia supernovae release an energy on the order of $10^{44}$~J~\cite{Branch1992}. 

If one uses the density $\rho=M/R^3$, the energy $E$, and the free-fall time $\tau=(G\rho)^{-\frac{1}{2}}$, then the radius of the Sun can be written as
\begin{equation}
R=\Big(\frac{E}{\rho}\Big)^\frac{1}{5} \tau^\frac{2}{5}\label{Taylor}
\end{equation}
This expression is reminiscent of the formula for the spreading of an explosion blast in the Taylor-Sedov regime~\cite{Taylor1950,Sedov1993}, which includes supernovae. Here, the free-fall time $\tau$ replaces the time since the beginning of the explosion. The size of the Sun is similar to that of an explosion frozen at a time $\tau$ after ignition. 

\subsection*{Equation of state and polytrope}
Note that Eq.~\ref{Geq1} should not be used to infer that $R\propto \Sigma^{-\frac{1}{4}}$, nor should Eq.~\ref{Geq2} be used to infer that $R\propto \Sigma^{\frac{1}{2}}$. Indeed, both mass and density can depend on pressure. The actual scaling between size and pressure will depend on the `equation of state' of the body; i.e. on the specification of an additional relation between $\Sigma$ and $\rho$~\cite{Landau2013}. We will see later that in some contexts this relation can be specified from microscopic considerations, but it can also be set phenomenologically by prescribing that the equation of state be that of a `polytrope'~\cite{Horedt2004}: 
\begin{equation}
\Sigma = K \rho^\gamma = \Sigma_r \Big(\frac{\rho}{\rho_r}\Big)^\gamma
\end{equation}
\noindent where in the first equation the dimensions of the proportionality factor $K$ are $\mathcal{M}^{1-\gamma}\mathcal{L}^{3\gamma-1}\mathcal{T}^{-2}$. The dimensionless exponent is usually written as $\gamma=(n+1)/n$, where $n$ is called the `polytrope index'. In the rightmost equation we define the proportionality factor {in terms of} a reference density and pressure as $K=\Sigma_r/\rho_r^\gamma$. The polytrope is expected to be valid in the vicinity of the reference values. 

Using Eq.~\ref{Geq2} in conjunction with the polytrope equation, we can express the radius $R$ as: 
\begin{equation}
R=\Big(\frac{K}{G}\Big)^\frac{1}{2} \rho^\frac{\gamma-2}{2} = \frac{\Sigma_r^\frac{1}{2}}{\rho G^\frac{1}{2}} \Big(\frac{\rho}{\rho_r}\Big)^\frac{\gamma}{2}
\label{polyR}
\end{equation} 
This approach is usually associated with Lane and Emden~\cite{Lane1870,Emden1907}. We will now see that values of $K$ and $\gamma$ can sometimes be computed from microscopic models.  

\subsection*{Microscopic density and pressure}
Whereas mass and energy are typically understood as `extensive' properties and depend on the size of the system, density and pressure are `intensive' and are expected to be independent of the size of the test volume. In fact, this independence of size only holds down to microscopic scales. Let us call $r$, $m$ and $\varepsilon$ the size, mass and energy of the smallest scale where $m/r^3=\rho$ and where $\varepsilon/r^3=\Sigma$. We call this scale the microscopic scale. For sizes smaller than $r$, density and/or pressure significantly vary from the macroscopic values. Using Eq.~\ref{Geq2} we can write: 
\begin{equation}
R^2=\frac{\varepsilon}{G m^2} r^3 = \frac{r^3}{d}\label{scales}
\end{equation} 
We know from Eq.~\ref{solarE} that $d=Gm^2/\varepsilon$ is the size of a body of mass $m$ at hydrostatic equilibrium under its own gravity and pressure $\varepsilon/r^3$. However, this size $d$ must be different from $r$. Indeed, by construction we have $r<R$ and $d<R$, and Eq.~\ref{scales} states that $R^2 d=r^3$, thus $d<r<R$. 

The identity between the microscopic and macroscopic pressures, $\varepsilon/r^3 = E/R^3$ is sometimes formulated as the `Virial theorem'~\cite{Clausius1870}: 
\begin{equation}
\varepsilon = G m \rho R^2 = G m \frac{M}{R} \label{Virialth}
\end{equation}
This equation relates the potential energy between the macroscopic mass $M$ and the microscopic mass $m$ to the microscopic energy $\varepsilon$. The energy $\varepsilon$ is usually understood as a kinetic energy involving the sound  speed $(\varepsilon/m)^\frac{1}{2}=(\Sigma/\rho)^\frac{1}{2}$.

\subsection*{Microscopic equilibrium}
In Eq.~\ref{scales} both lengths $d$ and $r$ are associated with the same energy $\varepsilon$ and mass $m$, but not in the same way. How can one express the size $r$? Can this be done in analogy with what we derived for the size of the Sun? Expressions such as $R=\Sigma/\Psi$ or $R=E/F$ are too vague to ever be false. They owe their generality to their vagueness. If one specifies the mass $M$, or the force $F$, or the energy $E$--or their densities ($\rho$, $\Psi$, $\Sigma$), then one can express more detailed results. This is what we did for the Sun, where the force comes from gravity. For the size $r$ we can similarly state without too much risk that: 
\begin{equation}
r=\frac{\varepsilon}{f}
\end{equation} 
\noindent where $f$ is a yet unspecified force. The only thing we know is that $f\neq Gm^2/r^2$; i.e. the dominant force at the microscopic scale is not self-gravity. Indeed, otherwise we would have $r=\varepsilon r^2 /Gm^2=d=R$ according to Eq.~\ref{scales}. Density and pressure would be far from intensive. If density and pressure are constant down to a scale of size $r<R$, then this scale must be governed by forces beyond gravity. 

\subsection*{Electromagnetic equilibrium}
The dominant force $f$ at scale $r$ can take many forms depending on contexts, but one example has proven to be very fruitful. This is the case where Newton's law of gravitation (\ref{Newtf}) is replaced by Coulomb's law of electrostatics (\ref{Coulf}): 
\begin{align}
F &= \frac{G M^2}{R^2}\label{Newtf}\\
f  &= \frac{k (ne)^2}{r^2}\label{Coulf}
\end{align}
For the gravitational force we assumed self-gravitation of a mass $M$ and size $R$, as in the case of the Sun. For the electrostatic force we assumed two opposite charges $ne$, where $e\simeq 1.6~10^{-19}$~C is the elementary charge and $n$ is an integer. The two charges are separated by a distance $r$. The Coulomb constant $k\simeq 9~10^9$~kg m$^3$ s$^{-2}$ C$^{-2}$ and the elementary charge can be combined into a single quantity $S_0=k e^2\simeq 2~10^{-28}$~kg m$^3$ s$^{-2}$ (no more units of charge). We then define $S=n^2 S_0$. The number $n$ is usually small due to screening between positive and negative charges. 

Using $f=S/r^2$ and $r=\varepsilon/f$ we can obtain the electromagnetic equivalent of the hydrostatic equilibrium:   
\begin{equation}
r=\frac{S}{\varepsilon} \label{QE} 
\end{equation}
In the same way that the energy $E$ and pressure $\Sigma$ were left unspecified in the gravitational case, here the nature of the energy $\varepsilon$ is unknown. For the gravitational case, we used the values of $M$ and $R$ to deduce the values of $E$ or $\Sigma$. Here, values of $r$ and $S$ would be enough to deduce a value of the energy $\varepsilon$ or pressure $\varepsilon/r^3$. 

\subsection*{Electromagnetic kinematics}
The substitution of the electromagnetic force in place of the gravitational force changes the definition of the kinematic quantities described so far. The formulas derived for the hydrostatic equilibrium of a body under its own gravity can be extended to the electromagnetic case by using the following substitutions: 
\begin{align}
g = \frac{GM}{R^2} & \rightarrow  \frac{S}{m r^2}\\
v_s^2 = \frac{\Sigma}{\rho} & \rightarrow \frac{\varepsilon}{m} \\
v_e^2 = \frac{GM}{R} & \rightarrow  \frac{S}{m r} \label{escapetrans}\\
\tau^2 = \frac{1}{G\rho} & \rightarrow \frac{m r^3}{S} \\
R_s = \frac{GM}{c^2} & \rightarrow  \frac{S}{m c^2}\label{EMRS}
\end{align}
For instance, in analogy with Eq.~\ref{gvs} one can say that $r=v_s^2/g$, only if the sound speed is redefined as $v_s\equiv( \varepsilon/m)^\frac{1}{2}$, and if the `electromagnetic acceleration' is redefined as $g\equiv S/(mr^2)$.

\subsection*{Stoney mass and Planck mass}
In the electromagnetic expressions, we have used the radius $r$, mass $m$ and energy $\varepsilon$, the same symbols we used in Eq.~\ref{scales} to describe the smallest scale at which density and pressure are equal to their macroscopic values. How can one be sure that electromagnetic forces dominate over gravity at the scale $r$? Using Eq.~\ref{scales} and Eq.~\ref{QE} we can write the size of the Sun {in terms of} the size obtained from electromagnetic equilibrium as:
\begin{equation}
R =\frac{(S G^{-1})^\frac{1}{2}}{m} r \label{sunEM}
\end{equation} 
This expression is legitimate if electromagnetism is the relevant force at the scale $r$. The strengths of the gravitational and electromagnetic forces can be expressed respectively as $Gm^2$ and $S$ (both with dimensions $\mathcal{M}\mathcal{L}^{3}\mathcal{T}^{-2}$). When $m=M$, the electromagnetic force is completely negligible since $GM^2 \gg S$ ($10^{50} \gg 10^{-28}$ kg m$^3$ s$^{-2}$, assuming $S=S_0$). One can then derive the mass $m_S$ that would correspond to the case where the electromagnetic and gravitational forces have similar strengths, i.e. when $R=r$:
\begin{equation}
m_S = \Big(\frac{S}{G}\Big)^\frac{1}{2} \label{StoneyMass}
\end{equation} 
In cases where the strength of the electromagnetic forces is well characterized by the value $S_0$ obtained for a pair of elementary charges, then $m_S \simeq 10^{-9}$~kg. This mass is called the Stoney mass~\cite{Tomilin1999}. Thus, Eq.~\ref{sunEM} can be rewritten as: 
\begin{equation}
R =\frac{m_S}{m} r \label{Stoney}
\end{equation} 
Similarly, we can write $d=(m/m_S)^2 r$. Note that if a material can locally deviate substantially from neutrality then the value of $S$ and $m_S$ must be increased accordingly.  

One can also define the fine structure constant as $\alpha=S_0/\hbar c\simeq 1/137$, with Planck's constant $\hbar\simeq 10^{-34}$kg m$^2$ s$^{-1}$ and reach: 
\begin{equation}
R =\alpha^\frac{1}{2} \frac{m_P}{m} r \label{Planck}
\end{equation} 
\noindent where the Planck mass is $m_P=m_S/\alpha^\frac{1}{2}=(\hbar c/G)^\frac{1}{2}$~\cite{Tomilin1999}. 

So far, we have used electromagnetism to derive a relation between the macroscopic $R$ and the microscopic $r$, but we have not specified any value for $r$ and $m$. From Eq.~\ref{QE} we know that specifying $\varepsilon$ will lead to $r$, but $m$ seems relatively free, as long as $m\ll m_S$. We know from Eq.~\ref{Stoney} that if the mass $m$ associated with the size $r$ is much smaller than $m_S$ then electromagnetism will be dominant over gravity. However other forces may dominate over electromagnetism, in which case Eq.~\ref{Stoney} or \ref{Planck} loose their validity. 

\subsection*{Atomic units}
A particularly interesting choice of microscopic size and mass considers the standard atomic units based on the hydrogen model with one nucleon (proton) of mass $m_0$ and one electron of mass $m_e$. This model is particularly useful for stars since they are mostly composed of hydrogen. In this context, we have $S=S_0$, $m=m_0\simeq 1.7~10^{-27}$~kg, and the Hartree energy and Bohr radius~\cite{Tomilin1999}: 
\begin{align}
\varepsilon_0 & = m_e \Big(\frac{S_0}{\hbar} \Big)^2 = \frac{\hbar^2}{m_e r_0^2} \simeq  4~10^{-18}~\text{J}\label{atomE}\\
r_0 & = \frac{S_0}{\varepsilon_0} = \frac{\hbar}{m_e \alpha c} \simeq  5~10^{-11}~\text{m} 
\end{align} 
\noindent where we used the electron mass $m_e\simeq 9~10^{-31}$~kg. Note that the ratio $S_0/\hbar$ has the dimensions of a speed, and is often expressed as $\alpha c$. With these units, pressure and density have values comparable to $\Sigma$ and $\rho$: 
\begin{align}
\Sigma_0=\frac{\varepsilon_0}{r_0^3} & \simeq  3~10^{13}~\text{Pa} \approx \Sigma\label{atomicSig} \\
\rho_0=\frac{m_0}{r_0^3} & \simeq  10^4~\text{kg/m}^3 \approx \rho \label{atomicden}
\end{align} 

With these atomic units we can obtain the following radius: 
\begin{equation}
R_0 =\frac{ \hbar^2}{(GS_0)^\frac{1}{2}} \frac{1}{m_0 m_e} =  \frac{m_P^2}{\alpha^\frac{1}{2} m_0 m_e} \ell_P   \simeq 5~10^7~\text{m}
\label{atomicR}
\end{equation}
In the last equation we express the reference size {in terms of} the Planck length $\ell_P=(\hbar G/c^3)^\frac{1}{2} \simeq 6~10^{-34}$~m, which is expected to be the smallest possible size. The radius is associated with the following mass: 
\begin{equation}
M_0 =\frac{m_S^3}{m_0^2} = \alpha^\frac{3}{2} \frac{m_P^3}{m_0^2}  \simeq 2~10^{27}~\text{kg}
\label{atomicM}
\end{equation}
These expressions only involve fundamental constants. We will say that $R_0$ and $M_0$ are the reference size and mass.  These types of expressions for the size and mass of stars are usually associated with the astrophysicist Chandrasekhar~\cite{Chandrasekhar1931}, {who} used these reference values in the derivation of the maximum mass of stable white dwarfs (now called the Chandrasekhar limit). The depth of these formulas has been noticed on several occasions~\cite{Weisskopf1975,Carr1979,Garfinkle2009,Burrows2014}. Note that Eq.~\ref{atomicM} can also be expressed, referring to Eq.~\ref{StoneyMass}, as an equation expressing the relative strengths of the electromagnetic and gravitational interactions: 
\begin{equation}
(GM_0^2)(Gm_0^2)^2 = S_0^3
\label{strengths}
\end{equation}

From the reference size and mass, one can also derive reference values for speed (sound/escape), acceleration and time (free-fall): 
\begin{align}
v_0 &= \Big(\frac{G M_0}{R_0}\Big)^\frac{1}{2} = \frac{\alpha}{\beta^\frac{1}{2}} c \simeq 5~10^4~\text{m/s} \label{refspeed}\\
g_0 &=  \frac{v_0^2}{R_0} = \frac{\alpha^4}{\beta^2} \frac{F_P}{M_0} =  \frac{\alpha^\frac{5}{2}}{\beta^2} \frac{F_P}{m_P}   \Big(\frac{m_0}{m_P}\Big)^2 \simeq 50~\text{m/s}^2\\
\tau_0 &= \frac{R_0}{v_0} = \Big(\frac{\beta}{\alpha}\Big)^\frac{3}{2} \Big(\frac{m_P}{m_0}\Big)^2 \tau_P \simeq 10^3~\text{s}
\end{align}
\noindent where $\beta=m_0/m_e \simeq 1836$ is the dimensionless ratio between the mass of a nucleon (proton or neutron) and the mass of an electron. The constant $\tau_P=(G\hbar/c^5)^\frac{1}{2}$ is the Planck time, and $F_p=c^4/G$ is the Planck force~\cite{Tomilin1999}.

\subsection*{From natural satellites to stars}
\begin{figure}
\centering
\includegraphics[width=8.5cm,clip]{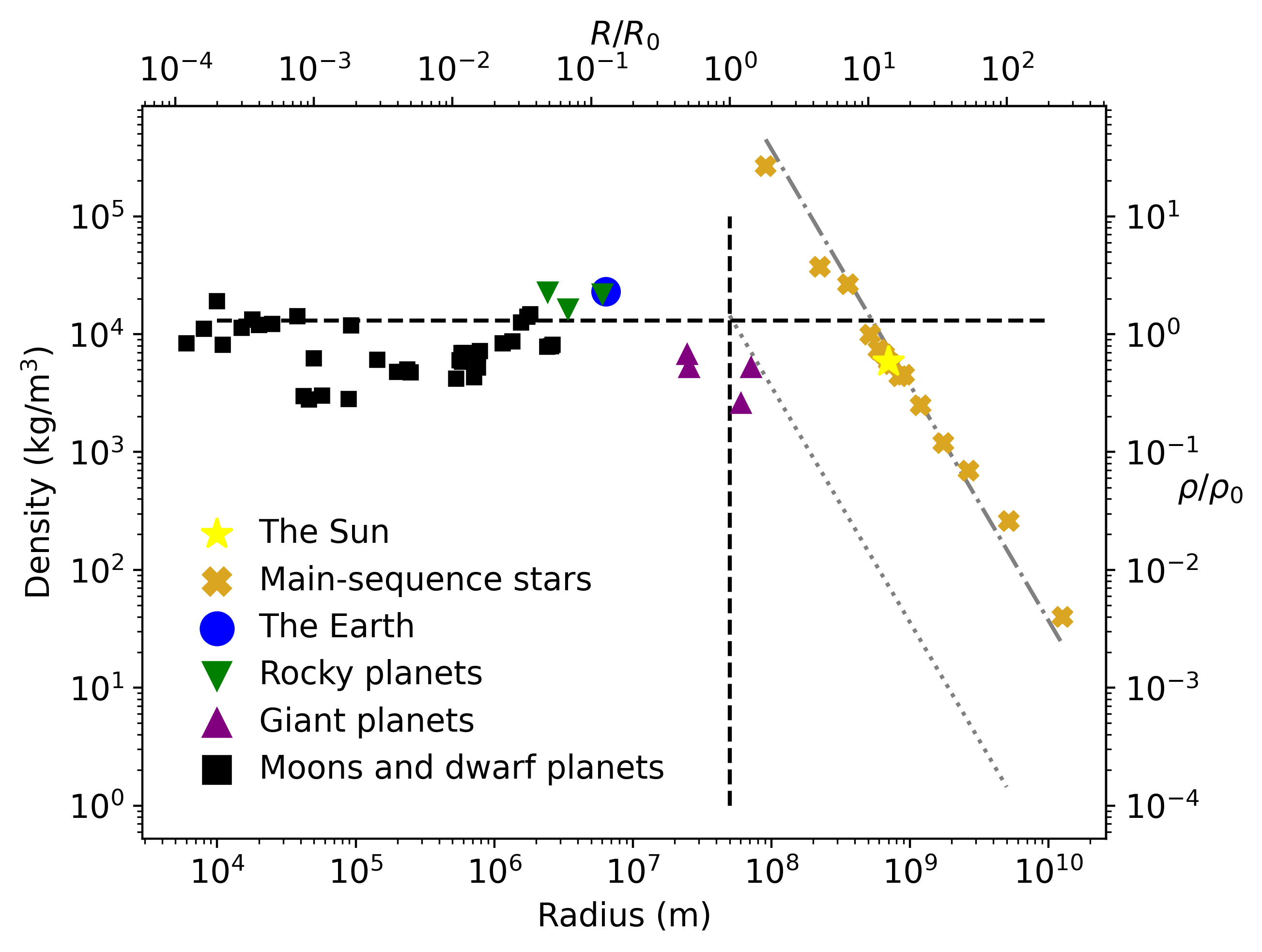}
\caption{The density as a function of radius for a few astronomical objects, from moons and planets to stars~\cite{Zombeck2006}. The vertical dashed line is $R_0$ defined in Eq.~\ref{atomicR}. The horizontal dashed line is the density $\rho_0$ obtained from Eq.~\ref{atomicden}. The dotted and dotted-dashed lines represent the predictions of the Virial theorem (Eq.~\ref{Virialth}), respectively with $\varepsilon=\varepsilon_0$ (Hartree energy), $\varepsilon=k_B T$ (thermal energy), with a constant temperature $T=10^7$~K.  Note that the density is here defined as $M/R^3$ instead of $M/(4\pi /3)R^3$, which gives values slightly larger than standard results.
\label{Fig1}}
\end{figure}

Given how we crudely neglected numerical factors, $R_0$ is not such a bad approximation for the radius of the Sun. However, there are stars with sizes smaller or larger than the Sun, for which the mass $m$ and energy $\varepsilon$ must differ from the atomic units. There can be atoms larger than Hydrogen, with $m>m_0$, and there could be atomic energies different from Hartree's formula. Deviations from the reference value computed {in terms of} atomic units can be revealed quite strikingly by plotting the density $M/R^3$ versus the radius $R$ for an array of astronomical bodies from natural satellites to stars~\cite{Zombeck2006}, as shown in Fig.~\ref{Fig1}. As can be seen the atomic units provide the size of the crossover between stars and planets, but stars, planets and smaller bodies can significantly deviate from the reference point. Moreover, stars display a different trend than smaller bodies.   

For stars, the density usually decreases as the radius increases. Whereas a star like the Sun has a density close to that of water, stars a hundred times larger can have a density smaller than that of air. This trend can be understood roughly by invoking the Virial theorem in Eq.~\ref{Virialth}, which states that $\rho=\varepsilon/GmR^2$. Thus, if the microscopic energy per unit mass $\varepsilon/m$ is constant then $\rho\propto R^{-2}$. In Fig.~\ref{Fig1}, the dotted line provides this scaling in the case where $\varepsilon=\varepsilon_0$, assuming $m=m_0$. As is obvious, this choice of atomic energy strongly underestimate the density. The dotted dashed line gives a better fit for an energy $\varepsilon=100 \varepsilon_0$. We shall explain such difference shortly. 

For planets and smaller bodies, the density is roughly constant, close to the reference atomic value $\rho_0$. For the density to remain constant for bodies of different sizes, the Virial theorem imposes that the kinetic energy per unit mass should follow $\varepsilon/m \propto R^2$. This can also be stated as $\varepsilon/r^3 \propto R^2$, or as $\Sigma \propto R^2$. The smaller the size the smaller the internal pressure $\Sigma$. 

The mass $m$, energy $\varepsilon$, and size $r$ of the microscopic scale can be fairly independent from one another. The microscopic parameters can produce different types of pressure $\varepsilon/r^3$ and density $m/r^3$, beyond the values obtained {in terms of} atomic units. In the microscopic realm, the relationship between $\varepsilon$, $m$ and $r$ is the equivalent of the macroscopic equation of state. In general, we can express the size $R$ in Eq.~\ref{Virialth} from the reference values as: 
\begin{equation}
R=\Big(\frac{\varepsilon}{\varepsilon_0}\Big)^\frac{1}{2} \Big(\frac{m}{m_0}\Big)^{-1} \Big(\frac{r}{r_0}\Big)^{\frac{3}{2}} R_0 =  \Big(\frac{\Sigma}{\Sigma_0}\Big)^{\frac{1}{2}}  \frac{\rho_0}{\rho} R_0
\label{Genref}
\end{equation} 
For stars or for planets, the particular relationships between the microscopic parameters can then be used to constrain this equation further. 

\subsection*{The Eddington model}
To give but one example of the ways in which the size of astronomical bodies can deviate from the reference value, let us consider the Eddington stellar model~\cite{Eddington1988}. This model is sometimes called `the standard stellar model', and it is usually described in textbooks concerned with stars~\cite{Kippenhahn1990}. Discussing this model also gives us the opportunity to underline the fact that our approach so far has been a bit stereotypical of an astro\textit{physical} viewpoint, in contrast to the astro\textit{nomical }perspective. Typical astronomical observations of stars rarely yield their size or mass, the most common observables being rather their luminosity and spectral characteristics. These observables are then transformed to mechanical quantities like mass, pressure, density, etc. The link for such a transformation is temperature.

In the Eddington model and others of its kind, the equation of state relating pressure and density uses the star temperature $T$. In practice, temperature varies inside stars from the core to the envelope, and $T$ must be understood as a bulk average. Using Boltzmann's constant $k_B\simeq 1.38~10^{-23}$~m$^2$ kg s$^{-2}$ K$^{-1}$, an energy can be translated into a temperature using $\varepsilon_T=k_B T$. In this framework Hartree's atomic energy would correspond to a temperature $T_0\simeq 3~10^5$~K. Thus, for the stars in Fig.~\ref{Fig1}, the energy $\varepsilon=100\varepsilon_0$ corresponds to a temperature $T\simeq 3~10^7$~K, which is the right order of magnitude for the temperature in stars like the Sun~\cite{Stix2012}. 

Eddington's model proposes that the pressure inside a star be given as a fixed combination of a {gas pressure} and a radiation pressure, where the {gas pressure} follows the ideal gas law $\Sigma_1=(\rho/m) \varepsilon_T$, and where the radiation pressure follows Stefan-Boltzmann law $\Sigma_2=\varepsilon_T^4/(c\hbar)^3$. If $\theta\in [0,1]$  is the proportion of thermodynamic pressure then $1-\theta$ is the proportion of radiation pressure, such that $\theta \Sigma =\Sigma_1$ and $(1-\theta) \Sigma =\Sigma_2$, where $\Sigma$ is the total pressure. One can then replace the thermal energy $\varepsilon$ by $m \theta \Sigma/\rho$ to express the radiation pressure as:
\begin{equation}
(1-\theta)\Sigma =\frac{ \Sigma^4}{(c\hbar)^3} \Big(\frac{m\theta}{\rho}\Big)^4  
\end{equation}
This equation can be solved for the total pressure $\Sigma$. By noticing that $c\hbar=S_0/\alpha$ the total pressure can then be expressed as: 
\begin{equation}
\Sigma = \frac{1}{\alpha}\Big( \frac{1-\theta}{\theta^4}\Big)^\frac{1}{3} \Big( \frac{m_0}{m}\Big)^\frac{4}{3} \Big( \frac{\rho}{\rho_0}\Big)^\frac{4}{3} \Sigma_0  
\end{equation}
If one assumes that $m=m_0$ and that $\theta$ is constant, then this equation provides a fully specified polytrope with $\gamma=\frac{4}{3}$ (and references $\rho_r=\rho_0$ and $\Sigma_r=((1-\theta)/\theta^4)^\frac{1}{3} \Sigma_0$). With this equation of state, the radius and mass of the star become: 
\begin{align}
R &= \frac{1}{\alpha^\frac{1}{2}}\Big( \frac{1-\theta}{\theta^4}\Big)^\frac{1}{6} \Big( \frac{\rho}{\rho_0}\Big)^{-\frac{1}{3}} R_0\\
M &=  \frac{1}{\alpha^\frac{3}{2}}\Big( \frac{1-\theta}{\theta^4}\Big)^\frac{1}{2} M_0
\end{align}
Stars following this equation of state can have different radii but they all share the same mass. If one assumes that the Sun follows such equation of state, it would imply that radiation only account for 20$\%$ of the pressure. Note that this equation of state implies $\rho\propto R^{-3}$. The data in Fig.~\ref{Fig1} are closer to $\rho\propto R^{-2}$, which can be interpreted by changing the proportion of radiation and thermodynamic pressures as the size increases. Arguments aimed at deciphering the value of $\theta$ for different populations of stars usually rely on a decomposition of the size of a star into layers (core, envelopes, etc.) with different amounts of radiation and convection. Schematically, greater accuracy on the size of stars requires their radius to be decomposed additively, for instance as $R=R_{core}+R_{env}$, where the sizes of the core and envelop are then decomposed multiplicatively as in the many ways we followed in this article~\cite{Kippenhahn1990}. This degree of precision goes beyond our scope but is inescapable if one wishes to recover the properties of the wide variety of stars in the cosmos.

\subsection*{White dwarfs, neutron stars and black holes}
\begin{table}
    \begin{tabular}{ | p{25mm}| p{25mm}| p{25mm} |}
    \hline
    \textbf{Object} & \textbf{Pressure} & \textbf{Radius} \\ \hline\hline
    White dwarf \newline(classical) & $\hbar^2/m_e r^5$\newline$\Sigma_0 (r_0/r)^5$\newline $\Sigma_0 (\rho/\rho_0)^\frac{5}{3}$ &~ \newline $R_0  (\rho_0/\rho)^\frac{1}{6}$   \\ \hline
    White dwarf\newline (relativistic) & $\hbar c/r^4$\newline$\Sigma_0 \alpha^{-1} (r_0/r)^4$\newline $\Sigma_0 \alpha^{-1} (\rho/\rho_0)^\frac{4}{3}$ & ~\newline $R_0  \alpha^{-\frac{1}{2}} (\rho_0/\rho)^\frac{1}{3}$   \\ \hline
   Neutron star \newline(ideal) & $\Sigma_0 \alpha^3 (\varepsilon_n/\varepsilon_0)^4$\newline$\Sigma_0 \alpha^{-1} (\rho_n/\rho_0)^\frac{4}{3}$ &  ~ \newline  $R_0 \alpha^\frac{1}{2} \beta^{-1} \simeq2$km  \\ \hline
 Black hole\newline (of mass $M_0$) & $c^8/G^3M_0^2$\newline$\Sigma_0 \alpha^{-4} (\rho_n/\rho_0)^\frac{4}{3}$  \newline $\Sigma_0 (\rho_b/\rho_0)^\frac{4}{3}$ &  ~ \newline  $R_0 \alpha^2 \beta^{-1} \simeq 1$m  \\  
    \hline
    \end{tabular}
\caption{Pressure and associated radius for different types of astronomical objects beyond the main sequence of stars. For a given object, the various expressions for the pressure are equivalent, assuming that the mass of the microscopic scale is $m=m_0$. The ideal neutron star considers a star with a nuclear density $\rho_n=m_0/r_n^3\simeq 10^{18}$~kg/m$^3$, where $r_n=\hbar/m_0 c$ is the Compton wavelength of the nucleon, and a nuclear energy $\varepsilon_n=m_0 c^2$. For the black hole of mass $M_0$, the density is $\rho_b=c^6/G^3 M_0^2$. We recall that $\beta=m_0/m_e \simeq 1836$ and $\alpha=S_0/\hbar c\simeq 1/137$. 
\label{Table1}}
\end{table}
All expressions of $R$ from Eq.~\ref{Geq1} to Eq.~\ref{Taylor} are essentially equivalent, they just present the same relations under different disguises. All these formulas correspond to an object bound by its own gravity, which is resisted by an unspecified internal pressure $\Sigma$, which can be expressed as an energy $E$ if need be. These expressions are expected to be valid for all objects in Fig.~\ref{Fig1}. Actually, these equations are expected to be valid for some objects beyond the range of the figure, in particular for stars outside the `main sequence'. 

The stars represented in Fig.~\ref{Fig1} all lie on the `main sequence', where the internal pressure is some mixture of thermodynamic or radiation pressures, well described by the Eddington model. However, some stars are governed by different types of internal pressure, which can lead to masses and radii lying outside the range of Fig.~\ref{Fig1}. Table~\ref{Table1} gives a few classical examples, left for the reader to ponder. 

\subsection*{Conclusion}
A diversity of `ideas for guessing' led us to more than a dozen formulas providing equivalent approximations of the size of the Sun. The size of the Sun can be seen as the result of a balance of forces or pressures, or as the consequence of an equation relating sound and escape speed, or as a frozen explosion. The size of the Sun can be expressed {in terms of} the size of its equivalent black hole, or {in terms of} {the size of its microscopic constituents}. We saw that some of these formulas also apply to other objects bound by gravity including different types of stars, as well as planets, and to some extent to microscopic objects bound by forces beyond gravity. 

The gigantic gap between the microscopic realm of atoms and the astronomical realm of planets and stars is quite daunting. In this paper, we provided a few simple arguments that can help bridge this divide, demonstrating the strong links between the quantum microcosm and the much larger scales dominated by gravity. 

The path we took to connect all these different formulas is one out of many. Any written story has a beginning and an end, so we had to start somewhere, with the hydrostatic equilibrium, and finish somewhere else, with the Eddington model and a table of additional cases based on different kinds of pressure. The beginning was in no way a means to an end, nor the end our goal. We invite the reader to explore their own path across the fascinating landscape of stellar physics, which provides an exciting laboratory for thought experiments about what we collectively mean by the `size' of something.

\textbf{Acknowledgments}: This paper was inspired by discussions with Nina Gnadig, Philip Avigan, Paul Cahen, Cyprien Gay and Vivek Sharma, and was fostered by the joyful atmosphere of the Ladoux-M\`{e}ge lab.

\end{document}